\documentclass[aps,prper,twocolumn,superscriptaddress
]{revtex4-2}
\usepackage[utf8]{inputenc}
\usepackage{enumitem}
\setlist{noitemsep}
\usepackage{tabularx}
\usepackage{makecell}
\usepackage{multirow}
\usepackage{longtable}
\usepackage{afterpage}
\usepackage{graphicx}
\usepackage{color}
\usepackage{ulem}

\usepackage[hidelinks]{hyperref}

\begin{document}

\preprint{APS/123-QED}

\title{Case Study: Context Interactions \& Physics Faculty’s Professional Development}

\author{Shams El-Adawy}
\email{shamseladawy@ksu.edu}
\affiliation{Kansas State University, Physics Department, 116 Cardwell Hall, 1288 N. 17th St. Manhattan, KS 66506-2601, USA}
\author{Tra Huynh}
\affiliation{University of Washington Bothell, 18115 Campus Way NE, Bothell, WA 9801}
\author{Mary Bridget Kustusch}
\affiliation{DePaul University, Department of Physics and Astrophysics, 2219 N. Kenmore Ave. Suite 211, Chicago, IL, 60614-3504, USA}
\author{Eleanor C. Sayre}
\affiliation{Kansas State University, Physics Department, 116 Cardwell Hall, 1288 N. 17th St. Manhattan, KS 66506-2601, USA}

\date{\today}

\begin{abstract}
This paper investigates the interactions between context and professional development of physics instructors in a case study of two physics faculty. A phenomenological-case study approach was used to analyze two physics faculty at different institutions over a year and a half using three semi-structured interviews each. The data enabled the identification of relevant context elements; and the impact of these elements on physics faculty was explored by adapting a framework that examines instructors' professional development. The analysis shows that both case study subjects used their physics expertise and growing understanding of their context to develop their physics teaching. However, this growth was enacted differently given the nature of their context, highlighting instructors' strengths in navigating their local context to improve their physics teaching. The results show the subtleties of how context has a salient, complex, and evolving role in moderating faculty’s professional development. By taking a faculty-centric approach, this paper broadens the community's awareness of the ways physics instructors develop their physics teaching. This work contributes to a relatively new lens by which the physics community views, discusses, and supports the professional development of physics faculty.
\end{abstract}

\maketitle
\section{Introduction}
This paper focuses on physics faculty professional development over time by examining the  role of context in moderating faculty's teaching practices.  Through the case study of two physics instructors, we focus on a faculty-centric discussion of physics faculty professional development by taking a longitudinal and holistic approach to examining their teaching.  We specifically  highlight the myriad of ways faculty bring together all their experiences to develop their physics teaching in order to answer the research question, \textit{how is the growth of physics faculty moderated by their context?}

Through various studies (described in \ref{subsection: faculty change work} and \ref{subsetion: role of context}), the literature has demonstrated that an instructor’s professional growth is an ongoing and dynamic product of factors relating to an instructor’s experiences in their environment. However, the literature has not significantly focused on physics faculty's strengths and agency in negotiating the interactions between the different entities that influence their professional development. Gaining a more detailed understanding of physics' faculty lived experiences as it relates to their practice provides an opportunity for the physics research community to enhance its continued support for faculty.  Drawing attention to all the nuanced ways faculty are improving their teaching can help the PER community enhance the ways it communicates research findings on faculty and their teaching to the broader physics community. Thus, this paper aims to broaden the community's awareness of the ways faculty develop their teaching by focusing on faculty agency in their context.   

\subsection{Faculty change work} \label{subsection: faculty change work}
Many research studies in discipline-based education research have been conducted on higher education instructors' adoption of research-based instructional strategies, the barriers they face in their implementation, and instructional transformation at the departmental and institutional level \cite{dancy2010pedagogical, dancy2016faculty, henderson2007barriers, henderson2011facilitating, henderson2012use, henderson2014assessment, turpen2016perceived}.  Research-based instructional strategies, which aim to promote student active engagement when learning physics, typically refer to named teaching practices at the core of many physics education studies. Examples of these instructional strategies are Just-in-Time Teaching, Peer Instruction, and Tutorials~\cite{henderson2009impact}. Studies have shown that instructors' knowledge of innovative teaching practices that are proven to improve student learning and retention does not necessarily translate into actual implementation of these strategies. Some research shows that instructors may discontinue using strategies after attempting to use them \cite{henderson2012use}  or they may greatly modify them in ways that render them ineffective \cite{henderson2007barriers}. This classical narrative of faculty change aims to encourage the use of established research-based instructional strategies and to better understand why faculty modify or discontinue the use of these methods \cite{henderson2012use}.

Building upon this classical narrative, lead physics education researchers on faculty development, Dancy and Henderson, worked on identifying the influence of both individual and situational characteristics that inhibit the implementation of PER based strategies \cite{henderson2007barriers}. Their work highlights the need for increased awareness of potential barriers \cite{henderson2007barriers}, the presence of a disconnect between instructors' conception of their practice and their actions in the classroom \cite{henderson2009impact} and the different forms of approaches to address instructional change \cite{henderson2011facilitating}. The latter particularly emphasizes how any type of lasting effective instructional change requires an in-depth understanding of an institution's complex system in order to design locally effective strategies \cite{henderson2011facilitating}.  Therefore, a framework for planning and implementing change at the institution level was developed \cite{henderson2011facilitating}. The framework creates what are referred to as \textit{departmental action teams}, which are interventions that bring together faculty, students and staff to enact departmental change. This holistic approach attempts to address some of the limitations of change in higher education, which include ignoring the complex interrelated nature of changing culture and university systems \cite{corbo2016framework}.

In parallel, studies have also highlighted the resourcefulness and need for ongoing support structures for physics faculty. The physics education community has and continues to develop programs to help physics faculty bridge the gap between research findings and physics teaching in their local context. Successful initiatives such as the New Faculty Workshop for Physics and Astronomy exist to encourage relatively new physics faculty to use research-based instructional strategies in their classrooms \cite{henderson2008physics}. These workshops have been proven to increase the use of research-based instruction, but physics faculty's use of these methods, as previously discussed, does not necessarily translate to a sustained implementation \cite{henderson2012use}. To address this continuity issue and to create ongoing support structures for faculty, programs such as a faculty online learning community (FOLC) model for educational change were developed~\cite{dancy2019faculty}. With the goal to support the transformation of faculty's teaching practices, these virtual communities create a support network for instructors to discuss, share, and help each other in their development of their teaching practices~\cite{dancy2019faculty}. This initiative is not the first to support physics faculty agency, but it is among some of the efforts in recent years that emphasize faculty's agency over their practice.
This avenue of research on faculty development aims to broaden the community's awareness of how physics faculty are viewed and discussed. In particular, this relatively new research lens is based on the premise that faculty's career experiences make them best suited to choose the research-based resources that best fit their own classrooms \cite{chasteen2020insights, dancy2016faculty, strubbe2020beyond, turpen2016perceived}.

\subsection{Role of context in faculty's professional development} \label{subsetion: role of context}
The role of local context has also been highlighted in the broader science community as crucial to support faculty's development of their teaching practices. Many studies in the sciences highlight the importance of local context for effective and lasting pedagogical interventions \cite{finelli2014bridging, gehrke2017roles, henderson2011facilitating}. The broader education literature also includes ways to assess the interpersonal dynamics that exist in context and their impact on the professional development of individuals.  The three components of Communities of Practice (the domain, the community and the practice) play a critical role in the ways in which an individual grows and flourishes within their field \cite{wenger2009communities}. Other education literature has highlighted the idea of situated teacher learning, which underlines how an instructor’s learning is intertwined with their classroom practices at their particular school \cite{putnam2000new}. Thus, teacher development needs to consider the interplay of these contextual factors \cite{putnam2000new}. Education literature has also put the spotlight on the interplay of different factors such as interpersonal relationships, institutional structures, personal considerations, and personal characteristics as supportive elements or impediments to the professional development of an instructor \cite{caffarella1999professional}. 

Furthermore, theoretical frameworks have emphasized the importance of overlapping contextual factors and their role on professional development of instructors. For instance, the Bell and Gilbert framework \cite{bell1994teacher} focuses on the contextual factors that come into play during a professional development program on teachers' growth. The Bell and Gilbert framework has been part of a range of existing models that account for the different ways teachers can improve their practice \cite{fraser33mckinney}. The model was created to assess professional development programs for science teachers in K-12 environments \cite{bell1994teacher} and puts forth three distinct, yet interrelated ways one can facilitate or restrict a teacher’s growth: personal (feelings, attitudes and beliefs about teaching developed through past experiences), social (working and relating to other teachers and students who are part of their community of practice) and professional (change of classroom practices such as using new activities and attempting new instructional approaches) \cite{bell1994teacher}. The creation of this framework showcased that teacher development is an on-going process that centers around each teacher’s unique journey to improving their teaching as they interact with professional development programs and other entities in their environment, environment, through which the theory illuminates ways one can facilitate or restrict a teacher's growth.  Concurrent with several education studies that examine the context-laden professional development of instructors \cite{mcchesney2019gets, voerman2015promoting}, the Bell and Gilbert framework provides a faculty-centric and stage based model for analyzing faculty’s growth, which is why we adapted it for our work  (more details on adaptation and implementation in Sec.\ref{sec: methodology} \& \ref{sec:theory}). 

\subsection{Summary of role of context and faculty change work}
As we can see from the above overviews of the literature, studies have emphasized 
the influence of contextual factors to instructors’ professional development in many ways. However, most research, particularly on physics faculty and their development, has  focused on the impact of particular programs or institutional structures on faculty’s development or on how to effectively help faculty adopt particular research-based instructional strategies. Research has not significantly focused on the interactions of multiple facets of the environment on faculty's professional development over time. Looking at a context-laden trajectory provides an opportunity to view and discuss faculty's change more holistically. The literature has not extensively examined the subtleties of how context has a salient, complex and evolving role in moderating physics faculty's professional development;  this paper aims to contribute to filling this gap.

\textit{Note on terminology}: We acknowledge that the use and meaning of professional development and development in the literature and in relation to our work has ambiguity and may be source of confusion. The literature often uses the idea of professional development interchangeably to refer to specific interventions or to broader facets of professional growth of faculty  \cite{ebert2011we,chasteen2020insights, olmstead2016assessing}. For clarity, in the rest of the paper, we use the term \textit{professional development programs} when discussing specific interventions and we simply use the term \textit{ professional development} to refer to all aspects of faculty's growth trajectory.
Moreover, with the exception of Fig.\ref{fig:bellgilbertframework} which was taken from Bell and Gilbert's original work, we  refer to the social development, personal development and professional development of the Bell and Gilbert's framework  \cite{bell1994teacher} as the \textit{social domain}, \textit{personal domain} and \textit{professional domain}, respectively. 

\section{Data}
The data discussed in this paper is part of a larger study aiming to improve the design of PhysPort, a website that provides instructional resources for physics educators  \cite{mckagan2020physport}. The study includes three sets of interviews with physics faculty at diverse types of institutions over the course of a year and a half, where each interview lasted about an hour over Zoom. Interviewees were asked to reflect on their existing and prospective teaching and assessment, needs, practices and philosophies as well as their use of student-centered teaching methods. 

The first set of semi-structured interviews was with twenty-three physics faculty at different types and sizes of institutions conducted at the beginning of the 2018-2019 academic year.  Eight of the twenty-three interviewees had follow-up interviews at the end of the 2018-2019 academic year and seven of the eight were interviewed again at the beginning of the 2019-2020 academic year. Therefore, we had access to three successive interviews for seven physics instructors. 

From our available interview corpus, Strubbe\textit{ et al.} used  three focal faculty for their work encouraging physics education researchers to think about faculty teaching using an asset-based agentic paradigm \cite{strubbe2020beyond}. For our paper, we chose two different case study subjects who were deemed good candidates because they were particularly articulate about their teaching practices and their environment. In addition, they both were relatively new physics faculty at their current institutions, but there was a clear contrast in their circumstances, which we deemed to be an interesting feature to explore in this type of project. We named our case study subjects Cleopatra and Sphinx. To explore their development as physics faculty, we adopted a phenomenological-case study approach with the aim of understanding their lived experiences. 

\section{Methodology}\label{sec: methodology}
The key features of phenomenology provided the tools to characterize the interactions between an instructor’s unique interactions with their environment, their personal experiences and their teaching. The focus of phenomenology is on the meaning or essence between participants and the world they experience through the study of a specific phenomenon \cite{dall2009phenomenology}. To understand the meaning of participants' lived experience, the methodology is based on three main steps: description, reduction, and interpretation \cite{sadala2002phenomenology, saevi2014phenomenology}. The description step consists of expressing the participant’s experience as they narrate it. The reduction step consists of a critical reflection by the researcher on the content of the description to identify emerging themes in relation to the project’s focus. The third step consists of interpreting the content identified in the description and reduction steps in order to give meaning to the participant’s story. Although the literature identifies a few different methods of analysis when using phenomenology, all approaches include these three general steps \cite{mapp2008understanding, earle2010phenomenology}.

In this paper, the phenomenon we are examining is the \textit{context-laden trajectory of physics faculty}. In practice, this is how the three-step method was conducted. First, the transcripts and video interviews of participants were viewed multiple times and a narrative description of the  interviews were written. Second, transcripts were re-read to locate significant statements or quotes about interactions between context and the faculty member’s development of their teaching practices. Quotes were regarded as relevant if they explicitly included information about practice and/or context. Third, to understand the essence of this phenomenon, emergent patterns within and across interviews for both context and development of faculty’s teaching practices were identified. The emergent patterns enabled a narrative description by the first author of what the instructor’s experience was and how they experienced it. Co-authors validated the narrative description by ensuring that the story accurately depicted the patterns identified in the data. Then, we connected relevant statements and themes to the broader literature about the phenomenon in order to interpret the meaning of participants’ lived experiences and implications for the phenomenon. Thus, after inductively identifying patterns, statements related to themes were clustered into the three domains of instructor development provided by the theoretical framework discussed in the following section (Sec.~\ref{sec:theory}).

A brief summary of the description phase can be found in Sec.~\ref{sec:description}. Then, Sec.~\ref{sec:claims} presents  a summary of the claims we are making. The justification for these claims from the reduction stage is presented in Sec.~\ref{sec:cleo-dis} and \ref{sec:sphinx-dis}, followed by the interpretation in Sec.~\ref{sec:interpretation}. Finally, we discuss the implications of work in Sec.~\ref{sec:implications}, followed by the limitations of this study and potential avenues for future work in Sec.~\ref{sec:limitations}. 

\section{Theory}\label{sec:theory}

To assess the ways in which faculty negotiate the interactions between the multiple facets of their context to develop their teaching, we adapted Bell and Gilbert’s framework  \cite{bell1994teacher}. The framework's three domains: personal, social and professional emerged from the assessment of a New Zealand professional development program where science teachers participated to enhance their teaching  \cite{bell1994teacher}. 

\begin{figure}[hptb]
  \includegraphics[width=\linewidth]{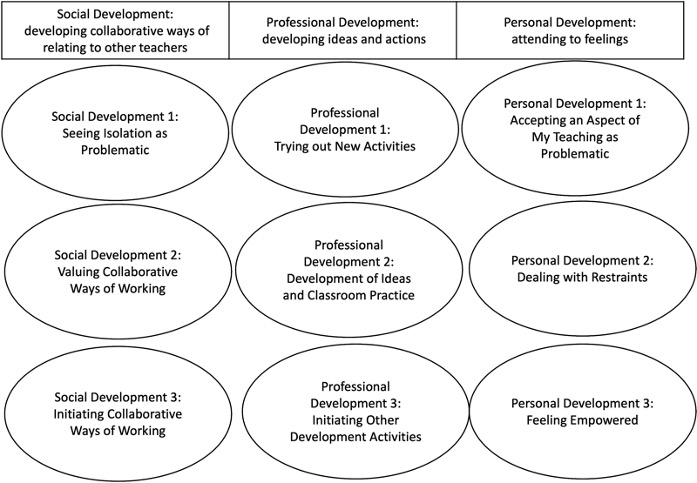}
  \caption{Bell and Gilbert’s framework’s diagram representation in the context of New Zealand’s science teachers’ development during a professional development program \cite{bell1994teacher} As indicated before, we will be using the term \textit{domain} to refer to each of these development spheres.}
  \label{fig:bellgilbertframework}
\end{figure}

These three domains were a result of the interview data collected to analyze this adult learning process in the context of improving teaching. Figure~\ref{fig:bellgilbertframework}, taken from Bell and Gilbert’s work, illustrates the overlap and interactions between these domains and their potential role in a teacher’s professional development. As the teachers went through the professional development program, each one of them went through phases of developing their teaching in each domain, as illustrated in Fig.~\ref{fig:bellgilbertframework} \cite{bell1994teacher}. Although each teacher had their own unique path  within these domains, the themes across each domain were of similar character for the participants in the study \cite{bell1994teacher}.

This framework highlights the idea that teachers' growth requires simultaneously fostering their growth in their professional, personal and social domains~\cite{chalmers2002exploring}. This framework puts the instructor at the center of their professional development by looking at the different ways these domains come together and evolve to moderate that instructor's growth. In particular, this framework centers the idea that the process of instructor's professional development is an interactive and interdependent undertaking in which social, personal and professional growth take place, with an emphasis that changes in one domain cannot happen without the other. Therefore, based on the themes that emerged from our data and the purpose of our study, we adapted this framework in order to have a faculty-centric approach to assess the different ways our case study subjects negotiate
their professional development as a result of the multitude of interdependent domains that play a role in moderating their physics teaching. 

However, we take the framework a step further. Whereas the framework is built to studying the impact of one professional development program, we use this framework to analyze the impact of several factors of the environment. We use the framework to analyze the impact of all the environmental factors and experiences faculty discuss that influence their growth as physics faculty over time. 
In other words, we extend the framework by considering the multifaceted, context-laden, trajectories of physics faculty as they develop their teaching practices. 

\section{Case study subjects}\label{sec:description}
\subsection{Descriptions}
Our first case study subject is Cleopatra, who is a tenure-track Assistant Professor of Physics at a military institution. She did her PhD after she retired from the military. She uses her hands-on experience in the military and her prior teaching experience outside physics to inform her physics teaching.  Cleopatra taught her third and fourth years at her current institution while taking part in these interviews, which spanned a year and a half. Throughout the interviews, Cleopatra emphasized how much she cares about her students and the way she adjusts and develops her physics teaching based on student understanding and her local context's constraints. 

Our second case study subject is Sphinx, who is a tenure-track Assistant Professor of Physics at a small liberal arts college. He uses his prior teaching experience as a graduate student teaching assistant to inform his physics teaching. He was in his second and third years of teaching at this institution while we conducted the three interviews. Throughout the interviews, Sphinx highlighted how much he cares about his students and the ways he is working on developing his physics teaching to improve student learning by adapting research-based methods to his context. 

 \subsection{Context Elements}\label{sec:context}
We characterized Cleopatra's and Sphinx's contexts in an organic way from the data based on their self-reported interactions with entities in their respective environments. 
Three levels of context were identified having an impact on their professional development: department, institution, and programs outside of their institution. Table \ref{tab:casestudy_context} summarizes the details of those contextual layers. Both Cleopatra and Sphinx have similarities in the members and experiences that make up their respective context such as colleagues inside and outside the department and their graduate school experience. However, some members and experiences are unique to their respective context such as the course director for Cleopatra and the college wide initiatives for Sphinx. The role played by these context entities will be discussed in the subsequent sections (Sec.~\ref{sec:cleo-dis} and \ref{sec:sphinx-dis}), as we analyze the interactions between context and their change in their social, personal and professional domains.
\begin{table*}[hptb]
    \centering
    \setlength{\tabcolsep}{5pt}
    \begin{tabularx}{0.95\textwidth}{p{1.5in}XX}
    \toprule
        \textbf{Level of Context} & \textbf{Cleopatra} & \textbf{Sphinx} \\ \colrule
      {Department}  & Students & Students \\ 
      & Colleagues & Colleagues \\ 
      & Course Colleagues & \\ 
      & Course Director & \\ 
      & Senior Colleagues & \\
      \colrule
      {Institution} & Colleagues in other departments & Colleagues in other departments \\ 
      & Regulations & College wide initiatives \\ 
      \colrule
      {Outside Programs} & New Faculty  Workshop for Physics and Astronomy  & New Faculty  Workshop for Physics and Astronomy \\
      & Graduate School Experience & Graduate School Experience \\ 
      & Past Professional Experience & \\
      \botrule
    \end{tabularx}
    \caption{ Elements of the three levels of context for Cleopatra and Sphinx}
    \label{tab:casestudy_context}
\end{table*}


 \section{Claims}\label{sec:claims}

For our subjects, we make the following claims about their professional development trajectory (summarized in Table \ref{tab:claims} for reference):
\begin{itemize}
    \item Cleopatra's growth as a physics faculty varies over time. Tensions with context members create agitation in her social domain, which echo in her personal and professional domains. Cleopatra's increased understanding of her context enables her to focus on what she can change in her teaching such as helping new instructors and focusing solely on her classroom practices (discussion in Sec.~\ref{sec:cleo-dis}).
    \item Sphinx's growth as a physics faculty is smooth. Harmony in his interactions with his context creates a smooth growth trajectory in his social, professional, and personal domains with the most significant growth occurring in his personal domain. Sphinx's increased understanding of his context allows him to realize that he has flexibility in what he can focus on day to day to keep developing his practice (discussion in Sec.~\ref{sec:sphinx-dis}). 
\end{itemize}
These claims show a contrast in the character of their professional development trajectory. However, for both, we claim that it was an increased context expertise that helped to facilitate this trajectory.


Interpretations of our claims were enhanced through our analysis in the Bell and Gilbert framework, which culminates in Fig.~\ref{fig:cleofigure} for Cleopatra and in Fig.~\ref{fig:sphinxfigure} for Sphinx.  As the developers of the framework established (Sec.~\ref{sec:theory}) and our data analysis suggests (Sec.~\ref{sec:cleo-dis} and \ref{sec:sphinx-dis}),  there is clear interconnectedness among the types of domains.

\begin{table*}[hptb]
    \centering
    \setlength{\tabcolsep}{5pt}
    \begin{tabularx}{0.95\textwidth}{X X}
    \toprule
        \textbf{Cleopatra} & \textbf{Sphinx} \\ \colrule
       Cleopatra's growth as a physics faculty varies over time. Tensions with context members create agitation in her social domain, which echo in her personal and professional domains. Cleopatra's increased understanding of her context enables her to focus on what she can change in her teaching such as helping new instructors and focusing solely on her classroom practices. (Sec.~\ref{sec:cleo-dis}). 
      & Sphinx's growth as a physics faculty is smooth. Harmony in his interactions with his context creates a smooth growth trajectory in his social, professional, and personal domains with the most significant growth occurring in his personal domain. Sphinx's increased understanding of his context allows him to realize that he has flexibility in what he can focus on day to day to keep developing his practice (Sec.~\ref{sec:sphinx-dis}).   \\
      \botrule
      \end{tabularx}
    \caption{Claims about Cleopatra's and Sphinx's growth}
    \label{tab:claims}
\end{table*}

\section{Cleopatra Discussion}\label{sec:cleo-dis}


The way Cleopatra interacts with her context impacts her three domains of development differently over time. 
Throughout her interviews, Cleopatra expresses an increasing frustration towards her restrictive context, which continuously presents challenges and barriers to her professional development. Concurrently, her better understanding of this unique context supports her by allowing her to refine her practice and mentor newer instructors. Her recognition of her more experienced local teaching status allows her to gain more agency in improving her practice.

\subsubsection{Interview 1} 
Central to Cleopatra's development is her desire to continuously improve her teaching, a significant feature of her personal domain. In tandem, different  interactions, more or less supportive, with members of her context moderate tensions in her social domain. These varied interactions, most notably in this interview with her students, course colleagues and senior colleagues, echo into her personal and professional domains. 

Her professional domain is centered around negotiating classroom practices, which includes using student-centered practices and finding ways to develop them. For example, one of the strategies she uses to help boost student participation is what she refers to as candy clicker questions, an instructional strategy  inspired by some colleagues in her institution.

 In addition to some of her institutional colleagues, one of the most influential context members to her professional domain is her students. Aligned with her student-centered approach, she attends to and prioritize her students' needs and interests in her teaching. For example, when engaging with students to contextualize physics problem, she draws on her past real experience in the military. This practice is responsive to her students' interest and at the same time showcases how her rich past professional experience informs her growth in the professional domain.
\begin{quote}
    \textit{``The kids are always asking for what they call war stories. Being retired military myself, I've got lots of stuff I can share, that's actually physics related.''}
\end{quote}
Moreover, professional development programs such as the New Faculty Workshop for Physics and Astronomy (NWF) provides resources that support her growth in professional domain. Specifically, Cleopatra draws ideas for her teaching strategies from NFW.
It is during NFW that she learned about the purpose of guided worksheets. Later, she uses NFW-inspired activities such as worksheets and ranking tasks to create her own set of conceptual multiple-choice questions.
Her interactions with the wider physics community enables her to put into perspective her teaching experiences and learn about new methods such as tutorials that she could potentially implement in her teaching.

In parallel, her personal domain in this stage is characterized by her continued openness to adapt her teaching to what best fits the circumstances of her context, which we refer to as an adaptive mindset. This mindset is moderated by supportive close colleagues and teachers in other departments who share her perspective on student-centered teaching and offer her insight on ways to continue to grow as an instructor. For example, she reaches out to department colleagues  to reflect on her teaching practices:
\begin{quote}
    \textit{``If I ever have a philosophy question about teaching I usually go and talk to [another female colleague in the department].''} 
\end{quote}

Simultaneously, her growth in the social domain is represented by how she is making efforts to enlarge collaborative ways of working with multiple members in her contexts. First, we notice that she discusses her interactions with students as a source of immediate feedback on how to continue to enhance her classroom practices. The level of engagement and the design of the course are elements of her teaching that are moderated by the students’ feedback that she values. For example, she incorporates instructional tools that students enjoy: 
\begin{quote}
    \textit{``The kids are more into the YouTube videos, Khan Academy, and stuff like that, and I give that to them all the time. I'm very heavy into the demos, and the internet videos, and then the PhET website, and stuff like that.''}
\end{quote}
Her increased focus on collaborating with her students strongly resonates with her increased orientation towards student-centered practices as shown earlier. In other words, we see that her growth in  her social domain strongly intertwines with her growth in her professional domain.

Second, she enlarges collaborative ways of working in her social domain by working with course colleagues. Cleopatra explains that instructors who teach the same introductory courses meet regularly to discuss the content of the class. As part of her community of practice, her course colleagues are the heart of her teaching environment and she attempts 
to collaborate with them when they meet bi-weekly to discuss course material. Cleopatra's mutual engagement with others is focused on helping her refine her ways of thinking as opposed to focusing on the development of those relationships themselves.

However, it is important to highlight that Cleopatra's overall growth is not supported by all members of her context. Her personal, social and professional domains are moderated differently by another set of people in her context--the higher level members within her department and institution. Cleopatra explains how the instructional design of the course is coming from higher level members in the department or the administration. There is a clear hierarchical structure to the department, where the course director and his assistant have full authority on creating, testing, and disseminating the in-class materials. This structure defines a limit to her agency on her practice.
\begin{quote}
    \textit{``Yeah, the syllabus is set by the course director.''}\\
    \textit{``[These new in-class activities] were vetted through the course director and the deputy course director.'' } 
\end{quote}

Additionally,  a certain group of people in Cleopatra's department are reluctant to change the way they teach: 
\begin{quote}
    \textit{``There’s ego involved there and a lot of [\dots] herding cats, I guess. People who are so used to doing it a certain way.''} 
\end{quote}
While her growth in the social domain centers around building collaborative ways of working, the institution and the department strongly encourages instructors to adopt and use materials as designed from the top down. This structure limits Cleopatra's growth in all domains.

Therefore, during the initial stage of this study, Cleopatra illustrates how she draws upon her experiences in outside programs to seek out interactions that provide insight to improve her teaching (characteristics of her professional \& personal domains), while finding ways to adapt to department rules and students’ needs as well as collaborate with other colleagues to continue to grow as an instructor (characteristics of her social \& personal domains). 

\subsubsection{Interview 2} 
During the second interview of this study, Cleopatra's growth in her social domain encourages her to advocate for changes in teaching practices in her classroom, which leads to growth in her professional domain and makes her feel empowered in her personal domain. During this phase, among the most salient members of her context are the course director and students. 

Along her trajectory within her professional domain, Cleopatra tries to find new best practices to implement, which is partially supported by the course director - a member who is part of the hierarchical structure of her institution that she described in her first interview. Hence, she feels empowered to make changes, a feature of her personal domain that is consistent with her desire to become a better instructor. Specifically, Cleopatra tries to engage with the course director to initiate different student-centered teaching practices and modify the use of some of their newest in-class activities:
\begin{quote}
    \textit{``I really wanted to do that [replace Project-Based Learning activities, PBLs] this semester, and I had kind of the approval of the course director to do it on a here-and-there kind of a basis, not ``no PBLs and you replace them with tutorials," but maybe one or two, maybe three, we could do that.''}
\end{quote}

 Cleopatra also discusses how her past experience with NFW continues to support her growth in her professional domain. In the following excerpt, she shares how she uses what she got from NFW to supplement her teaching and create good clicker questions for students:
\begin{quote}
    \textit{ ``I was borrowing pieces [from ranking tasks book from AAPT workshop] when I wanted a good clicker question, a good solid one, I would flip through the book and go, ‘Oh yeah, I like that, but I'm going to change it in this way and add it to my slides and have it as a clicker'.''}
\end{quote}


In parallel, within her social domain, she reiterates the idea of continuing to collaborate with course colleagues, which showcases the central role of these constant and regular interactions to her trajectory in her social domain and her desire to continue negotiate classroom practices, a continuous characteristic of her professional domain. 

She also continues to value informal feedback from students to tailor her instructional methods in the best way possible for them: 
\begin{quote}
    \textit{``I took the opportunity to actually ask my students after the first day I got back, I said,  `I have a unique opportunity to get feedback from you guys' and then the two questions were, `I like it better when Dr. Cleopatra\dots.' `I like it better when the substitute teacher\dots.' Then I could get, `Oh, we really like it when you do this, but boy, they really did a great job doing that.' ''}
\end{quote}

Furthermore, based on students’ feedback and preferences, she makes adjustments to her instructional design strategies and attempts new ways to structure her classroom time with students. For example, since students wanted to engage in different types of activities during class time, Cleopatra decided to switch up more frequently throughout the class period between whole class discussion of the content and small group activities. 

Nevertheless, some of the tensions persist within the department between members who are reluctant to new ideas and those who are not, which restrict what she can and cannot do within the social domain: 
\begin{quote}
    \textit{``I’m trying to keep the big picture here, but we've got a lot of academy grads here that say, `well, that’s not the way I went through', \dots so it’s frustrating when you work with somebody like that. `oh, we got to go back to the way it was.' No, we don't.''}
\end{quote}
This persistent tension leads her to focus on aspects she can change in her social domain. The narrowing down to aspects of teaching she can change allows her to continue feeling empowered in her teaching. Thus, she tries to focus on parts of her position where she has the most agency and impact such as helping newer instructors adjust: 
\begin{quote}
     \textit{``People like me with experience, then, can offer it to the new guys, and a lot of times, they’ll actually come and watch us teach.''}
\end{quote}
Her attempt to enlarge collaborative ways of working (characteristic of her social domain) gives her a sense of empowerment (characteristic of her personal domain). Her stronger agency over her teaching and her place at her institution helps keep finding ways to implement new and better practices  (characteristics of her professional domain), despite some of these persistent constraints that she understands more with time. 

\subsubsection{Interview 3}
In interview 3, Cleopatra faces greater tensions and constraints from members of her social domain, particularly the course director, who restrain her possibilities of growth in her professional  domain. Thus, she narrows her focus to  continue improving her teaching within the constraints of her context in her personal domain with the support of colleagues in other departments and the wider physics community.

The accumulation of strict regulations  puts more boundaries in what she can do in her continuous search for best practices, a main feature of her professional domain. For instance, new enforced departmental rules inhibit her from continuing to use her candy clicker questions, an instructional strategy that used to increase student engagement in the classroom: 
\begin{quote}
    \textit{``So what's happened this semester, which is very sad is that the department head and the upper ups have locked down the food in the classroom. So I can't even have the candies in the classroom, I would have to wait to give it to them like on the way out the door or something.''}
\end{quote}

This interview marks a significant growth in Cleopatra's understanding of her local context. In the following excerpt, she highlights how her increased understanding over time allowed her to identify one of the the institutional mechanisms that make her context restrictive and upholds the power dynamics at play: 
\begin{quote}
    \textit{ ``Yeah, and I do volunteer for this position [course director] but it doesn't seem like any civilians are getting picked for these jobs. So I'm assuming anyways at this point that they're keeping it military for course directors only for some reason.''}
\end{quote}

It is not surprising that course directors, higher-level members in departments and institutions are pivotal in Cleopatra’s context because the structure of institutions of higher education gives them vital power to shape and shift practices. More specifically, Cleopatra is at a military institution, which has even more peculiarities than other institutions. Previous studies have shown that departmental and institutional structures are decisive factors in instructional change in physics \cite{corbo2016framework}.  Nevertheless, the interesting and noteworthy highlight in Cleopatra’s case is how she decides to navigate this environment. Cleopatra turns her focus toward the context members and practices that allow her to continue to grow as an instructor across the three domains.

In her personal domain, she starts to accept the constraints of her local context. 
There is a clear structure and power in place that inhibit instructors from making changes to certain in-class material, despite the self-reported ineffectiveness of some the teaching materials: 
\begin{quote}
    \textit{``We talked about it [the pros and cons of discovery projects] all the time in our group meeting. So all in all the teachers that teach the same course, will meet and talk about it. [\dots] so people aren't loving it. Instructors aren't loving it. Students aren't loving it. But somebody in charge says let's keep doing it.''}
\end{quote}

Additionally, in her personal domain, she moves towards finding ways to deal with the constraints of her restrictive situation and she focuses on what she has the most power to develop, which is her student-centered in-class practices: 
 \begin{quote}
     \textit{``I am still not seeing so much benefit. It's, um, it's a little crazy. I feel like just when I'm getting them into the board work, and we're, we're ready to go into the next harder problem. I go, Oh, I gotta stop and we got to go do this discovery exercise [activities during which students encounter new physics concepts and/or topics]. And a lot of times I feel like the material doesn't actually line up with what we're doing very well. So it almost feels like busy work to me. I'm still trying to reconcile that. I'm trying to figure out how do I make them better? What could I do?  How could I help make them better? Well, how would I change them and stuff and I'm usually just barely grading them on time and then I don't even think about them anymore.''}
 \end{quote}

Alongside her thinking about ways to deal with the restraints, she starts to recognize her own context-dependent expertise and her role as a resource to help newer instructors develop their teaching within the same local context. 
Although she cannot be in a position of power such as course director, she finds ways to have a supportive leadership influence by helping newer instructors navigate the terrain of this restrictive institution: 
\begin{quote}
    \textit{``One of the two new instructors who just showed up this semester is shadowing me, and watching my class to see how the labs go and how the discovery exercises go.''}
\end{quote}

Her better understanding of her restrictive context allows her to mentor newer instructors. 
Simultaneously, in her social domain, she focuses on what she can change by continuing to engage with supportive resources in the wider physics community  to keep growing as an instructor. For instance, she reiterates in her third interview how the wider physics community gave her the idea to implement tutorials in her classroom to help scaffold student learning of new concepts.

Furthermore, the negogiating of current classroom practices is accompanied by her search to find ways to implement new practices that fit her local context. She continues to engage with students to get informal feedback about their preferred methods of teaching and learning. For example, she considers implementing a new instructional strategy students wanted because they enjoy using in other courses. The strategy called “the wheel of fortune” where the instructor turns the wheel with student names on and calls upon the student the wheel stops at it to answer the question:
\begin{quote}
    \textit{``[Interviewer: do you think you should use this new teaching idea because it's a really great idea or because it would just be fun?] I think both actually, because the students that I was talking to, I got some informal feedback from them [\dots] that this would be pretty cool to implement.''}
\end{quote}

By finding new strategies for classroom engagement, Cleopatra continues to find ways to implement new practices and negotiate classroom practices, taking into account students' input and resources from the wider physics community. 

In this interview, we see that Cleopatra focuses her attention on finding ways to continue to engage in improving student learning and engaging with supportive colleagues (characteristic of her social domain).  As her understanding of the constraints of her context settles (characteristic of her personal domain), Cleopatra narrows her focus to her classroom, where she can enact her agency and continue to improve her teaching within her broader restrictive context (characteristic of her professional domain).

\subsection{Cleopatra Summary}\label{sec:cleo-sum}
 \begin{figure*}[tb]
  \includegraphics[width=0.7\linewidth]{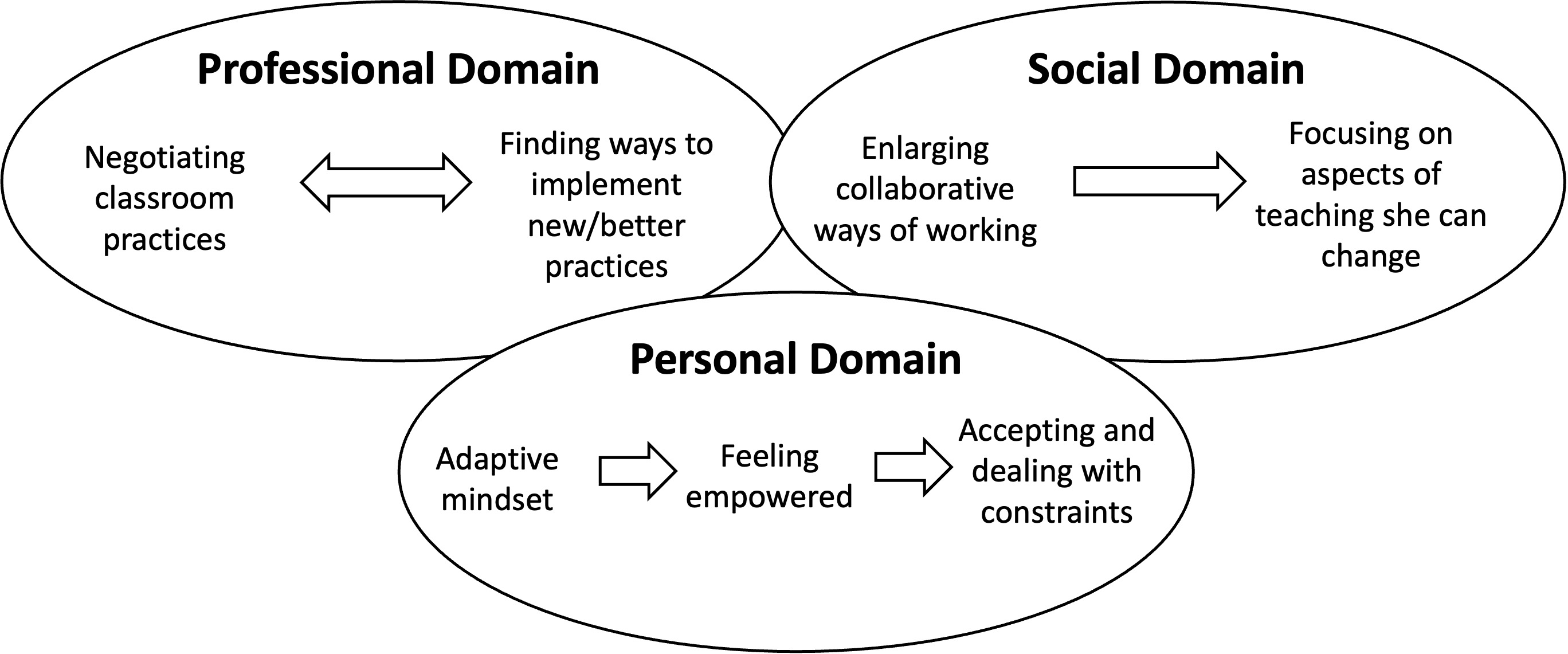}
  \caption{ Cleopatra’s Trajectory Characterization in the Bell and Gilbert Framework}
  \label{fig:cleofigure}
\end{figure*}

Figure~\ref{fig:cleofigure} summarizes the key features of Cleopatra’s context-laden trajectory within the Bell and Gilbert's framework. In the professional  domain, Cleopatra engages in negotiating between teaching practices and the constraints of the system in order to find ways to implement new or better practices. In the social domain, she starts by attempting to enlarge her collaborative ways of working, then she focuses on what she has the most power to change. In her personal domain, she starts with an adaptive mindset, which leads her to feel empowered in her ability to change her practices, but then she grasps the limits of what she can do and attempts to deal with the constraints due to the nature of her context. 

Our analysis of subtle changes in the way Cleopatra discusses her practice and her context across interviews suggests an overarching pattern that indicates a growing context-expertise in developing her physics teaching, but her growth as an instructor is tumultuous due to her context. Her recognition of her more experienced local teaching status allows her to gain more agency in improving her practice. However, through interactions with key actors in her context, she better understands the unique restrictions of her institution and she focuses more deeply on ways to develop the instruction of introductory physics that she can control. 

\section{Sphinx Discussion}\label{sec:sphinx-dis}


The way Sphinx interacts with his context impacts the three domains of development similarly over time. Of his three domains, his growth is most noticeable in his personal domain, which echoes through his social and professional domains. 
\subsubsection{Interview 1} 
During the initial phase of the study, Sphinx feels that he has conflicting ideas about the best way to teach. Nevertheless  he discusses exploring activities and practices in order to figure out what might work in his environment. His growth in his social domain is strongly nurtured by his institution and department colleagues. 

In his personal domain, Sphinx strives to continuously improve his teaching. During this exploratory phase, he reflects on what he can do to improve and align with his current understanding of best practices from the physics education research community. He describes his perception of how knowledge of physics is constructed as partially through transmission from the instructor to the students and he emphasizes the need for brief lecturing:
\begin{quote}
    \textit{``I'm still a little bit in the camp where I think there should be a little bit of lecturing. I don't think it has no place in the classroom.''} 
\end{quote}
Sphinx seems to believe this approach to teaching contradicts what the physics research community promotes as best practices.
However, his discussion of his teaching practices reveals that his practices are largely what the community would describe as student-centered. 
For example, he uses formal and informal feedback from students to better understand them and refine his teaching accordingly: 
\begin{quote}
    \textit{``I wanna be fluid, I wanna respond to them. I'll even ask them for feedback at certain points, after exams and whatnot. I've looked at their surveys and what it is they want.''} 
\end{quote}
Sphinx's current openness to adapt his teaching, which we refer to as an adaptive mindset in his personal domain, stems back to an experience in his graduate studies, where he was exposed to ways to negotiate
classroom practices to implement best practices.
In particular, he was exposed to ideas around active learning during a workshop as a teaching assistant in graduate school. Sphinx explains that the biggest takeaways from this workshop were reflected in his current grading in his classes. He now implements frequent low risk assessments instead of a few high risk assessments. 

Concurrently, Sphinx enlarges his collaborative ways of working in his social domain. His approach and thinking about teaching evolves as he interacts with colleagues outside the department about ways to improve his pedagogy, where he seems to seek out interactions that help develop his teaching. For example, he participates in a campus wide reading group where he meets regularly with other faculty members from campus to discuss common readings about best teaching practices. 

In addition, Sphinx is in a small department and discusses regularly with the other two faculty how things are going and how to make changes to improve their current teaching methods. 
\begin{quote}
    \textit{ ``We talk about things like this a little bit in department meetings. It's a small department. There's only three full-time faculty. We talk about these things a little bit. How is this year's cohort doing? How's everything going?''}
\end{quote}

Furthermore, Sphinx explains how he uses some of his college wide initiatives to shape his approach to teaching. He gets involved in activities initiated by the Center for Excellence in Teaching and Learning (CETL) at his institution. For example, during this phase of our study, he had participated in workshops organized by CETL:
\begin{quote}
    \textit{``They have other things they do, different workshops. We have a workshop to start the fall semester where we talk about teaching practices and mentoring and advising practices. They sent out an email and I thought it would be good to dive into that.''}
\end{quote}
Sphinx's expanding flexible context enables him to decide what to focus on at any given time in order to keep developing.  Sphinx’s context-expertise translates into his ability to develop his physics teaching through implementation of incremental changes. 
His context is growing, dynamic and flexible, which allows him to easily try and implement little changes in his practice, which demonstrates growth in all three domains.

In this initial stage, we found that Sphinx interestingly presents a different case from findings of previous work. Henderson \cite{henderson2009impact} highlighted the disconnect between instructors' conception of their practice, which often happened to favor student-centered approaches, and their actions in the classroom, which were often more instructor-centered practices. However, Sphinx’s conception and practice align with student-centered strategies, but his perception is that his conception of his practice is not in agreement with what the research on student-centered approaches in the classrooms suggests. 

Despite Sphinx's perception that his teaching is in conflict with best research-based teaching practices, Sphinx seeks out ways to continue to improve his practice (characteristics of his personal domain). His discussion of his practice suggests a strong emphasis on student-centered teaching and learning (characteristic of his professional domain). His interactions are facilitated by increased and supportive social interactions in his environment (characteristic of his social domain).

\subsubsection{Interview 2}
Over time, Sphinx interacts with new members of his context such as the wider physics community, which nurtures his growth in his social domain. As a consequence, these interactions open new possibilities  for his own conception of teaching, feature of his personal domain, and new possibilities for his classroom practices, feature of his professional domain. His context members already part of his local network also continue to give him flexibility in what he can do. 

During the second phase of this study, Sphinx enlarges his collaborative ways of working in multiple ways. First, his recent interactions with members of the wider physics community at NFW is at the forefront of his discussion. He highlights the ways in which the workshop benefits his practice:
\begin{quote}
    \textit{``After the workshop happened, I used them [PhET simulations] a couple times in the fall, but I definitely used them more in the spring. I thought that was a good one, it's a nice visual. Some of them are good visual representations.''} 
\end{quote}
Moreover, after attending NFW, he immediately incorporated PhET simulations in his course:
\begin{quote}
    \textit{``After the workshop happened, I used them [PhET simulations] a couple times in the fall''} 
\end{quote}
In his small department he has the flexibility in implementing new strategies as he sees fit.

Second, Sphinx easily discusses with his department colleagues how his teaching is going and finding new ways to continue to improve practices. For example, he is making labs more student-centered with one of his colleagues in the department. 
\begin{quote}
    \textit{``The other thing was how we structure the labs. Our labs at the moment, well, I guess we're already in the midst of making a change. There are two new faculty members in my department, so the other one went as well to the workshop, and we both kind of were talking about some of the stuff we learned from it. And then, in the department meetings, we're a small department, so the two of us, we're two-thirds of the department, and then we came back and said, well, these are things we would like to maybe change, and we sort of gauged receptiveness to it.''}
\end{quote}

We can see that his enlargement of collaborative ways of working echoes into his growth in his professional domain, where he strives to find better ways of teaching. While working closely with his colleagues in restructuring their labs,  Sphinx continues to reflect upon his experience in graduate school to negotiate how to best improve his classroom practices. In graduate school, Sphinx was a lab instructor and a recitation instructor. Reflecting upon his approach to these teaching assignments, he realized how the one-sided approach he had adopted then did not create an engaging environment for learning for his students. He would show students how to use the equipment in his pre-lab lecture and answer questions during the lab. In recitation, he was talking and working through the problems, without many interactions with students. His recent interactions with the wider physics community and with members of his own college made him realize how this teaching strategy was unhelpful to students because
\begin{quote}
    \textit{``It was really one-sided where they didn't get to talk so much.''} 
\end{quote}
Additionally, this one-sided approach did not allow him to adapt to students' needs and let them have agency over their own learning of the material. Thus, in his current faculty position, he is making sure to engage with his students.
In fact,  some of the interactions he values are conversations with students on ways to better tailor classroom practices to students' needs. For instance, he discusses how getting feedback from students is an important aspect of his assessment and development of his teaching: 
\begin{quote}
    \textit{``In talking with some of the Physics majors that are seniors, some of them think like, I'd rather have it be more interactive, go through less things but in depth more, rather than cover every single chapter.''} 
\end{quote}

Simultaneously, we learn that one of the reasons he decided to work at this small liberal arts institution was the opportunity to easily engage with colleagues and students: \begin{quote}
    \textit{``I chose to apply to a place like that because I wanted to work closely with people, with students. There's a small student to faculty ratio and then also I could do some summer work with them.''}
\end{quote}

In interview 2, thanks to new and continued interactions with his environment (characteristics of his social domain), Sphinx reflects on his trajectory as an instructor and tries new teaching strategies that favor student-centered practices and looks for ways to keep improving his practice (characteristics of his personal \& professional domain). 

\subsubsection{Interview 3} 
During interview 3, Sphinx focuses his attention back to particular aspects of his teaching as he is being evaluated for promotion, while continuing to be nurtured by his supportive context members.

Within his professional domain, Sphinx continues to find new ways of teaching, while he enlarges collaborative ways working, the consistent feature of his social domain. It seems that outside programs such as NFW and his graduate school experience continue to easily integrate into Sphinx's discussion of all facets of his teaching. As previously mentioned, a teaching workshop during graduate school exposed him to the importance of student engagement for learning.
Then, recently, attending NFW brought back this idea to the forefront of his teaching and developed it even further.  In fact, Sphinx acknowledges the pivotal role  his interactions with the wider physics community at the workshop played in his conception, assessment, and discussion of his physics teaching: 
\begin{quote}
    \textit{``I would say it [the New Faculty Workshop for Physics and Astronomy] definitely influenced how I teach. So when I was in graduate school, I had taken some kind of  teaching workshop or something for graduate students and so there they had already kind of sort of impressed upon me maybe the importance of getting the students to be more active. But I think that was definitely brought home at the New  Faculty Workshop for Physics and Astronomy.''}
\end{quote}

Additionally, Sphinx's department as a whole is trying to improve their teaching practices, which means Sphinx finds himself in a department that encourages and values  constantly finding ways to develop one's teaching, which creates alignment between the department and Sphinx in terms of professional domain growth. He continues to support  one of his colleagues who is in the process of rewriting some of the introductory labs to make them more inquiry based. He is increasingly appreciating that his small department easily offers the opportunity to develop and implement new and better teaching practices. 

Given that Sphinx was being evaluated for promotion during interview 3, we see a shift in his personal domain from discussion of big picture ideas of teaching coming from his adaptive mindset to  specific instructional strategies. His discussion on his teaching focuses significantly on the day to day classroom activities:
\begin{quote}
    \textit{``I have a candidate committee who reviews me in my class, my classes. I am also this year getting evaluated college wide, so I had someone from the college as well evaluate [\dots]''}
\end{quote}

More specifically, he discusses in detail strategies he uses to engage students in the classroom. For example, he uses ABCD cards (cards with the letters A B C and D that allow students to respond to questions an instructor asks in class) to include more conceptual questions to gauge student understanding and increase their engagement: 
\begin{quote}
    \textit{``And so, for those conceptual questions, that's where I use the ABCD cards. And so I'll have them you know, I use them for pre questions, so basically before I go over the topic just to see where everybody's at. And so when it looks like everybody's on the same page. Like, I've tended to go a little quicker through those topics. If ever there's kind of a mixed response then like, I usually hold off on telling them the answer then we go through the material and then I ask them again.''}
\end{quote}
His discussion on his in-class teaching strategies showcases how he tries to adapt to students' level of understanding and create a pedagogical structure that centers around student interactions. 

Additionally, he continues to have an adaptive mindset, which continues to be a feature his growth in his personal domain. 
We observe this feature when he details his thought process behind his use of certain instructional methods such as how he implements incremental changes to provide more opportunity for peer discussion: 
\begin{quote}
    \textit{``If I see differences in answers, like I'll try to get either two people to talk to each other or to talk through me about what they're thinking and then or sometimes I'll have them talk to each other as well and read, discuss their answers.''} 
\end{quote}
The value of peer instruction is a facet of his teaching that is strongly nurtured by his supportive environment. His personal past experiences, his interactions with the wider physics community, his college and his colleagues all strongly encourage this mode of instruction.
Later on, he tries to make adjustments as he gets feedback from colleagues from outside his department: 
\begin{quote}
     \textit{``This semester a colleague [...] said, to make sure you know, when you call on people, you know, try not to always call on the same people [...] try to like distribute it out a little more.''}
\end{quote}

His student-centered in-class practices are now more robust thanks to the tools and strategies he gained by enlarging his collaborative ways of working and engaging in finding new and better practices.


As we can see, in interview 3, Sphinx shifts  his attention to day to day practices (characteristics of his personal domain) but continues to engage in collaborative ways of working  with students and colleagues (characteristic of his social domain). Negotiating classroom practices (characteristic of his professional domain) is facilitated by the inherent flexibility of his type of institution and his discussion also suggests that his work is strongly nurtured by a supportive environment. 

\subsection{Sphinx Summary}\label{sec:sphinx-sum}

\begin{figure*}[tb]
  \includegraphics[width=0.7\linewidth]{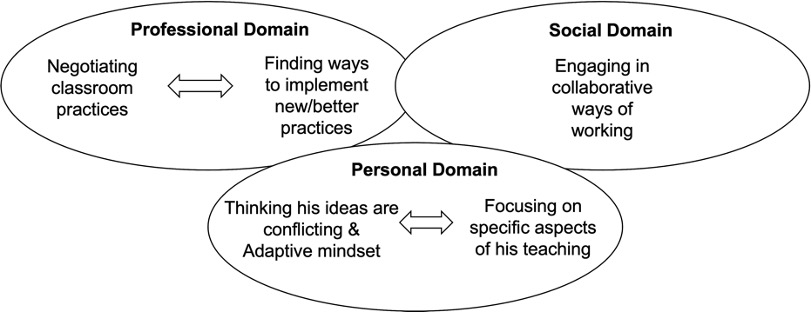}
  \caption{ Sphinx’s Trajectory Characterization in the Bell and Gilbert Framework}
  \label{fig:sphinxfigure}
\end{figure*}
 
Figure~\ref{fig:sphinxfigure} summarizes Sphinx’s smooth trajectory within the Bell and Gilbert framework. In the professional domain, Sphinx engages in negotiating between classroom practices in order to find ways to implement new or better practices. In the social domain, he increasingly engages in collaborative ways of working. In his personal domain, he starts with the perception of a conflicting mindset, which through interactions with his context shifts to his openness in embracing new ways of teaching, but then he focuses on particular facets of his teaching. 

 Our analysis of how  Sphinx discusses his practice and his context among interviews suggests an overarching pattern that indicates a growing context-expertise and growing confidence in his ability to develop his physics teaching through implementation of incremental changes. His interactions with his context over time allow him to gain agency in improving his practice. The flexibility of his context enables him to have a smooth professional development trajectory. Through his interactions with the wider physics community and his colleagues, he engages and takes more concrete actions about ways to restructure introductory physics courses in his small department. His personal, social and professional domains are being incrementally aligned with his approach to enhance his teaching, which facilitates his ongoing professional development as a physics instructor.

\section{Interpretation}\label{sec:interpretation}
Instructors' unique experiences undoubtedly leads to the development of their physics practices. Nevertheless, our case study subjects show some of the interesting nuances that exist in the interactions between context and the development of their physics teaching. Both Cleopatra and Sphinx decide to focus on developing their largely student-centered in-class practices. However, the reasons they focus on these practices are different because of the nature of their distinct contexts, which creates the opportunity to examine the interesting nuances that moderate faculty change. For Cleopatra, her experience in her local context enables her to navigate tensions and focus on the aspects of her teaching she is able to develop. For Sphinx, his experience in his local context enables him to coordinate his teaching practices to best fit his current needs such as the current assessment of his teaching by his college.

A closer look at both case study subjects' characterization in the Bell and Gilbert framework (Figs. \ref{fig:cleofigure} and \ref{fig:sphinxfigure}, respectively) shows growth in their professional domains by finding ways to implement new and better practices, yet this development is moderated differently by their social and personal domains. Sphinx's environment fosters collaborative ways of working, which is the central aspect of his social domain over time, whereas Cleopatra's efforts to enlarge collaborative ways of her working are not as suited for her restrictive context, which leads her to narrow down the areas of collaborative work in her social domain over time. Moreover, in their respective personal domains, Cleopatra and Sphinx have contrasting trajectories. Cleopatra's trajectory goes from flexibility and empowerment to accepting and dealing with the constraints of her context, whereas Sphinx's development is directed towards an adaptive mindset focusing on particular aspects of his daily teaching.
These context-laden characterizations in our extension of the Bell and Gilbert framework illustrate two ways physics faculty develop over time. 

 Understanding the experiences of these two physics instructors gives perspective to physics education researchers about the ways physics faculty negotiate their myriad experiences to continue to  develop as physics faculty. In particular, it highlights the benefits of the  lens by which we conducted the analysis of this study and adds to what previous work on faculty change has done.  
 
 First, our study extends on the classical narrative of faculty change which aims to assess adoption and persistence of use of established research-based strategies \cite{dancy2010pedagogical, dancy2016faculty, henderson2007barriers, henderson2011facilitating, henderson2012use, henderson2014assessment, turpen2016perceived} in instructional change by providing a more holistic and faculty-centric perspective of both Cleopatra and Sphinx's stories. Although the classical narrative would have captured some of the research-based strategies our case study subjects use, its intended purpose  does not allow us to capture how faculty moderate the use of these strategies in light of their interactions with different facets of their context. 
 
 Second, we analyzed the effects of a combination of experiences our case study subjects underwent, which provides more nuances on factors that can impact physics faculty's professional development. Assessing the impact of a particular intervention such as NFW would have captured quite well Sphinx's professional and social domain, but would not have as clearly characterized his growth over time. In fact, NFW played an incredibly positive role for Sphinx who discussed its benefits at length in our study. This is concurrent with recent assessment studies of NFW's positive impact on faculty \cite{chasteen2020insights}.
 In contrast, assessing the impact of a particular intervention would not have been as helpful  for Cleopatra. While she does greatly benefit from programs such as NFW, a more crucial layer of Cleopatra's story is the significant role played by her institution and how she is learning to navigate the tensions of her context to continuously grow as a physics instructor. Previous studies and frameworks were not created to examine faculty change work with the lens we did in our analysis.
 
 In fact, by using the Bell and Gilbert framework \cite{bell1994teacher} we were able to use similar theoretical grounds for both case study subjects that highlight similarities and differences in both instructors' professional development. The framework was valuable for analyzing faculty growth as it was used to look at the different aspects of their context that can impact their professional development. Then, our extension of the framework enabled us to show the nuances and the multifaceted factors that come into play in moderating faculty change over time.
By studying faculty growth from this lens, we attempted to look at faculty development more holistically, offering more reasons to find more nuanced ways for the physics education research community to view, discuss and support physics faculty. 

\section{Implications}\label{sec:implications}
Although studies in physics education research have shown the importance of context in instructional change and the need for continued support for faculty while implementing research-based methods \cite{dancy2010pedagogical, dancy2016faculty, henderson2008physics, henderson2011facilitating,henderson2012use, henderson2014assessment, turpen2016perceived}, they have not focused on the varied, context dependent expertise instructors bring to their physics classrooms, and they have not always taken a faculty-centric approach to physics faculty’s development. Therefore, by highlighting two different possible ways physics faculty's growth is enacted in their local context, we contribute to a current trend in physics education focusing on asset-based models that support the different ways physics faculty navigate their professional development in their local context \cite{strubbe2020beyond}.
Shifting the way we view and discuss physics faculty by taking this faculty-centric approach enables a more holistic and tailored analysis of their experience, which creates the possibility of creating more ways to help moderate continued support for physics faculty.

First, our study underlined the important role of colleagues within a local context. Both Cleopatra and Sphinx engaged with either department colleagues and/or institution colleagues to continue to enhance their practices. Local colleagues are an integral part of our case study subjects' growth trajectory as they help them develop their context expertise. In addition, in Cleopatra's case, she then offers that mentorship to newer colleagues as they join her department. Department and institution colleagues, especially for early career instructors, play a critical role in helping instructors' navigate teaching in their specific context. Therefore, the PER community should not only encourage and support engagement of instructors with the wider physics community, but also engagement with supportive local colleagues.

Second, our study highlighted one example of a conflicting message the PER community sometimes inadvertently sends to non-PER faculty about what active learning entails. Messages such as \textit{lecture is  bad} can be seen, such as by Sphinx, to suggest that his teaching approach did not align with his understanding of what PER championed as best practices for teaching. PER can do a better job at communicating what they mean by active learning, because the perception is that there is not any time given to presentations or lectures, even brief ones, by the instructor.  Our communication with the broader physics community often does not articulate well that  many successful physics classrooms that incorporate active-learning are coherently structured with a combination of lecture and student-centered activities.

Finally, as physics education researchers continue to develop resources and find ways to best disseminate their findings to the broader community of physics educators, they have the opportunity to highlight  two concrete ways junior physics faculty navigate very different contexts to develop their teaching. Cleopatra and Sphinx's stories are two examples of how the process of becoming a physics faculty member is a multifaceted and ongoing journey, which could be used in professional development programs such as NFW. The contexts of our case study subjects provide concrete examples of how junior faculty utilize explicit and implicit interactions with their surroundings and experiences to develop their physics teaching. This case study could be used to help other junior faculty navigate this stage of their career. Through the contrasting examples of Cleopatra and Sphinx, other junior physics faculty have two different, yet successful, examples of how to navigate their unique circumstances to develop their physics teaching.

\section{Limitations \& Future Work}\label{sec:limitations}
Our analysis is based on self-reported data by our case study participants. On the one hand, this limitation enabled us to do an in-depth phenomenological-case study  to narrate the complexity of the trajectory of faculty's lived experience. On the other hand, the identified patterns and interplay of faculty's lived experience cannot be generalized across all physics faculty. Although this feature is an inherent part of any case study, this limitation makes the purpose of our study scope broadening one, where we broaden the community's awareness of the ways physics instructors develop their teaching. 

Moreover, as one might expect, our case study subjects' institutions played an evident role in their contrasting experiences. The inherent size and type of institution clearly had a significant role in moderating their practice. The deliberate choice we made in choosing two participants with different environments contributes to the purpose of the paper, which is to show the complex and evolving role of context in faculty's professional development. Yet, our analysis shows that context-laden faculty professional development goes beyond the type of environment  they find themselves in, which highlights the value of a faculty-centric approach to better understand the nuances of their lived experience to best support them.

In the future, given our case study subjects' illustration of the different, complex, and context dependent expertise instructors bring to their classrooms to help student learn physics, we hope to expand this work to more physics instructors in different environments to enlarge the discussion of physics faculty professional development, focusing on the strengths of their context-dependent teaching, beyond their physics, or even pedagogical knowledge.

In conclusion, our study illustrates the growth of physics faculty, specifically non-PER faculty. Our methodological approach showed how Cleopatra and Sphinx's physics faculty experiences enabled them to utilize their communities' influences and their physics expertise to navigate their physics teaching to best fit their local circumstances. The subtleties of how context has a salient, complex, and evolving role in moderating faculty’s professional development can continue to contribute to a more faculty-centric approach in the ways the physics community views, discusses and supports the professional development of physics faculty.

\begin{acknowledgments}
We wish to thank the physics faculty who participated in this study. We also wish to thank the PhysPort research team for access to the data collected and the KSUPER group for their support. Additionally, this paper grew from the first author's master's thesis so we would like to thank DePaul University's physics department, particularly members of the first author's thesis committee for their support. Lastly, we would like to thank the reviewers for their helpful feedback on earlier versions of this manuscript. This project is based on work supported in part by the following funding sources: DePaul Summer Research Grant, DePaul Graduate Research Fund, a PERLOC (Physics Education Research Leadership and Organizing Council) Travel Grant and NSF 1726479/1726113.
\end{acknowledgments}

\bibliography{references.bib}

\end{document}